\documentclass[pss]{wiley2sp} 
\usepackage{bm} 

\usepackage{float}
\usepackage{pst-node}
\usepackage{amsmath}
\usepackage{graphicx}
\usepackage{latexsym}
\usepackage{amsfonts}
\usepackage{amssymb}
\usepackage{graphicx}
\usepackage{rotating}
\usepackage{bbm}
\usepackage{cancel}
\usepackage[nouppercase]{scrpage2}
\usepackage[T1]{fontenc}

\newcommand{\be}{\begin{equation}}
\newcommand{\ee}{\end{equation}}
\newcommand{\bea}{\begin{eqnarray}}
\newcommand{\eea}{\end{eqnarray}}

\def\a{\alpha}
\def\b{\beta}
\def\g{\gamma}

\def\d{\delta}

\def\th{\theta}

\def\L{\Lambda}
\def\m{\mu}
\def\n{\nu}

\def\r{\rho}
\def\s{\sigma}
\def\S{\Sigma}
\def\t{\tau}

\def\vf{\varphi}

\def\blr{{\mathbf r}}

\def\callL{\mbox{$\mathcal{L}$}}

\def\callS{\mbox{$\mathcal{S}$}}

\def\1op{\hat{\mathbbm{1}}}
\def\nn{\nonumber}

\begin{document}

\title{The Dissection Algorithm for the second-Born 
self-energy}

\author{%
  Enrico Perfetto\textsuperscript{\textsf{\bfseries 1}} and
  Gianluca Stefanucci\textsuperscript{\Ast,\textsf{\bfseries 
  2},\textsf{\bfseries 3}}
 }

\mail{e-mail
  \textsf{gianluca.stefanucci@roma2.infn.it}}

\institute{%
  \textsuperscript{1}\,CNR-ISM, Division of Ultrafast Processes in Materials
(FLASHit), Area della ricerca di Roma 1, Monterotondo Scalo, Italy\\
  \textsuperscript{2}\,Dipartimento di Fisica, Universit\`a di Roma Tor
Vergata, Via della Ricerca Scientifica, 00133 Rome, Italy\\
  \textsuperscript{3}\,INFN, Sezione di Roma Tor Vergata, Via della Ricerca
Scientifica 1, 00133 Roma, Italy}


\abstract{\bf%
We describe an algorithm to efficiently compute  
the second-Born self-energy of many-body perurbation theory. 
The core idea consists in 
dissecting the set of all four-index Coulomb integrals into properly chosen 
subsets, thus avoiding to loop over those indices for which the Coulomb 
integrals are zero or negligible. The scaling properties of the 
algorithm with the number of basis functions is discussed. The 
computational gain is demonstrated in the case of 
one-particle Kohn-Sham basis for 
organic molecules.
}

\maketitle   

\section{Introduction}

At the beginning of the new century the computational capabilities 
were powerful 
enough to
implement the nonlinear integro-differential Kadanoff-Baym equations 
(KBE) put forward in the sixties~\cite{kadanoff1962quantum,danielewicz1984quantum,svl-book,balzer2012nonequilibrium}. 
The first KBE calculation dates back to 2000 and deals with plasma 
oscillations in the homogeneous electron gas~\cite{PhysRevLett.84.1768}. 
KBE calculations of inhomogeneous systems like atoms and diatomic 
molecules driven out of equilibrium by external laser fields appeared in 
2007~\cite{PhysRevLett.98.153004}.  Since then the number 
of groups working on efficient implementations of the KBE has grown, 
and the interest has progressively
moved toward nonequilibrium properties of model 
Hamiltonians. Among the pioneering works we mention
Refs.~\cite{mssvl.2008,mssvl.2009,pva.2009,pva.2010,PhysRevB.93.054303,PhysRevB.81.115131,PhysRevA.82.033427,BalzerHermanns2012,Sakkinen-2012}.

Despite the aforementioned advances, the KBE have not yet been 
combined with ab initio schemes for the investigation of  
nonequilibrium properties of realistic systems. In fact, the propagation of a two-times object like 
the Green's function is still too burdensome, even for most modern 
supercomputers. An enormous simplification to the numerical solution 
of the KBE 
occurs when evaluating the collision integral using the Generalized 
Kadanoff-Baym Ansatz (GKBA)~\cite{PhysRevB.34.6933}. For this reason 
there has been a considerable effort in assessing the 
reliability of the 
GKBA~~\cite{HermannsPRB2014,LPUvLS.2014,CTPPBonitz2016,C60paper2018} and 
in combining it with ab initio 
methods~\cite{Pal2011,PSMS.2015,PUvLS.2015,Sangalli-2016,Pogna.2016,SangalliEPL2015,PSMS.2016,PSMS.2018}
over the last years.
Through the GKBA the nonequilibrium Green's function formalism is converted into a 
time-dependent density-matrix functional 
theory~\cite{PhysRevA.75.012506,Giesbertz2010,Giesbertz2014} with the advantage that correlation 
effects can be included through diagrammatic approximation to the self-energy.

As pointed out in a recent work~\cite{KvLPS.2018}, the GKBA is an 
ansatz for the Green's function and the computational gain with 
respect to a full KBE simulation is limited to 
self-energies up to the second Born (2B) level, with first- 
and second-order exchange diagrams evaluated using either the bare 
Coulomb interaction $v$ or the statically or partially dynamically 
screened interaction $W$. The 2B approximation
well reproduces equilibrium spectral properties~\cite{schuler2017} 
and total energies~\cite{DahlenLeeuwen2005} of molecular systems.
Furthermore,  benchmarks against numerically 
exact simulations in 1D atoms and molecules~\cite{PhysRevA.82.033427}, 
quantum wells~\cite{BalzerHermanns2012}, weakly correlated 
Hubbard and extended Hubbard
nanoclusters~\cite{Sakkinen-2012,HermannsPRB2014,CTPPBonitz2016,HopjanPRL2016,ReichmanEPL2016,Joost2017},
the Anderson model at finite bias~\cite{UKSSKvLG.2011} and 
photo-excited donor-acceptor tight-binding 
Hamiltonians~\cite{C60paper2018} indicate that the 2B approximation 
remains accurate even out of equilibrium.
The GKBA implementation of more sophisticated 
approximations like the self-consistent GW 
or T-matrix self-energies would scale like the original KBE unless a GKBA-like expression for the fully dynamically 
screened interaction $W$ or T-matrix $T$ is provided.

The special role played by the 2B diagrams (with $v$ or $W$ 
interaction lines) in the GKBA scheme has triggered the search for efficient 
algorithms to compute the 2B self-energy. In an electronic system with $N$ 
one-particle degrees of freedom the computational cost of a naive implementation would scale like $N^{5}$. 
Such an unfavourable scaling 
makes first principles  simulations of 
systems with more than two or three light atoms numerically expensive.
In this work we describe a {\em dissection algorithm} which takes the maximum 
advantage of vanishing or very small Coulomb integrals. The 
dissection algorithm is currently implemented in the CHEERS 
code~\cite{PS-cheers} and it allows for first principles simulations of 
systems with tens of active electrons like, e.g., organic molecules, 
up to tens of femtoseconds~\cite{PSMS.2018}. We emphasize that the 
algorithm does not make any use of the GKBA and, therefore, it can 
also be implemented   
to simulate time-dependent quantum phenomena within the full KBE 
dynamics.

The paper is organized as follows. In Section~\ref{sigmasec} we 
introduce the equation of motion for the one-particle density matrix 
and write down the explicit 
form of the 2B self-energy. The dissection algorithm is described in 
detail in Section~\ref{algorithmsec}. In Section~\ref{scalingsec} we 
discuss how the computation of the 2B self-energy scales with the 
number of basis functions. Extensions and generalization of the 
algorithms along with an outlook on future applications is provided 
in Section~\ref{conclusionsec}.

\section{The 2B self-energy}
\label{sigmasec}

We consider a finite system of interacting electrons described by $N$ 
one-particle basis functions $\{\vf_{i}(\blr)\}$. For simplicity  we assume that the 
Hamiltonian is invariant under spin rotations and that we have an 
equal number of up and down spin electrons (the results derived in 
this work do not rely on these assumptions and can easily 
be generalized). The one-particle density matrix $\r$ is diagonal in 
spin space and the matrix element $\r_{ij}$ between $\vf_{i}$ and 
$\vf_{j}$ is independent of the spin orientation.
The equation of motion for the $N\times N$
matrix $\r$  
reads~\cite{LPUvLS.2014,PUvLS.2015,C60paper2018,PSMS.2018}
\be
\frac{d}{dt} 
\r_{ij}(t)+i[h_{\mathrm{HF}}(t),\r(t)]_{ij}=-I_{ij}(t)+\mathrm{H.c.} \, ,
\label{gkba}
\ee
where $h_{\mathrm{HF}}(t)$ is the time-dependent HF Hamiltonian and 
$I(t)$ is the collision integral, written in terms of the greater and 
lesser self-energy and Green's function~\cite{svl-book}
\be
I(t)=\!
\int_{0}^{t}\! d\bar{t}\!\left[\S^{>}(t,\bar{t})G^{<}(\bar{t},t)-
\S^{<}(t,\bar{t})G^{>}(\bar{t},t)\right].
\label{collint-for-rho}
\ee
In the 2B approximation  the self-energy 
$\S$ reads
\be
\S_{ij}=
\sum_{\substack{n\\ pq\\ sr}}
v_{npri}G_{pq}\bar{G}_{sr}
\sum_{m}
\left[2v_{mqsj}G_{nm}-v_{mqjs}G_{nm}\right],
\label{2Bself}
\ee
where we used the short-hand notation
\be
\S\equiv\S^{\lessgtr}(t,t') \quad\;\quad G\equiv 
G^{\lessgtr}(t,t')\quad\;\quad\bar{G}\equiv G^{\gtrless}(t',t).
\ee
For real one-particle basis the Coulomb integrals in 
Eq.~(\ref{2Bself}) are given by
\be
v_{ijmn}\equiv \int d\blr d\blr'
\frac{\vf_{i}(\blr)\vf_{j}(\blr')\vf_{m}(\blr')\vf_{n}(\blr)}{|\blr-\blr'|},
\ee
and have the symmetry properties
\be
v_{ijmn}=v_{jinm}=v_{njmi}=v_{imjn}.
\label{symprop}
\ee

The computational cost for evaluating $\S$ for a given pair of times $t,t'$  depends on how 
much the 
tensor of Coulomb integrals $v_{ijmn}$ is sparse. Recalling that $N$ 
is the 
dimension of the one-particle basis,
the cost can vary from $N^{2}$ for a Hubbard-like interaction, i.e.,
$v_{ijmn}\propto \d_{ij}\d_{im}\d_{in}$, to $N^{5}$ when all 
$v_{ijmn}$'s are nonvanishing.
In the latter case, however, several Coulomb 
integrals may be order of magnitude smaller than 
others~\cite{PUvLS.2015,PSMS.2018},
and the calculation of the r.h.s. of Eq.~(\ref{2Bself})  
remains accurate by considering only those $v_{ijmn}$
larger than a certain cutoff $\L$. 
In fact, the 2B self-energy is 
quadratic in $v$ and therefore $\L$ can generally be chosen 
larger than the cutoff of a typical Hartree-Fock (HF) calculation (the HF 
self-energy is linear 
in $v$). Therefore, the effective scaling with $N$ can be 
considerably reduced if one manages to sum over only 
those indices for which $|v_{ijmn}|>\L$.
The basic idea consists in dissecting the set of all Coulomb integrals
in properly chosen subsets.

\section{Dissection algorithm}
\label{algorithmsec}

In this Section we illustrate an efficient algorithm to calculate
the r.h.s. of Eq.~(\ref{2Bself}) with $v_{ijmn}\to 
v_{ijmn}\th(|v_{ijmn}|-\L)$.
We begin by noticing that the first 
term in the square bracket is nonvanishing only if the pair $(qs)$ 
belongs to 
\be
\callL=\{(qs):|v_{mqsj}|>\L\;{\rm for\;some}\;(mj)\},
\ee
and that the second term in the square bracket is nonvanishing only if the pair $(qs)$ 
belongs to 
\be
\callL_{\rm x}=\{(qs):|v_{mqjs}|>\L\;{\rm for\;some}\;(mj)\}
\ee
Let $D$ and $D_{\rm x}$ be the dimension of the sets $\callL$ 
and $\callL_{\rm x}$ respectively. 
Given $\a=1,\ldots D$ we establish a map to 
extract the couple $q=q(\a)$ and $s=s(\a)$. Similarly, given 
$\b=1,\ldots,D_{\rm x}$ we establish a map to 
extract the couple $q=q(\b)$ and $s=s(\b)$. Once these maps are 
defined we introduce a superindex $I=1,\ldots N\times D$ from which 
to extract $n={\rm Int}[I/D]+1$ varying between $1$ and 
$N$ and $\a=I-(n-1)D$ varying from $1$ to $D$. 
Similarly we introduce a superindex $J=1,\ldots N\times D_{\rm x}$ from which 
to extract $n={\rm Int}[J/D_{\rm x}]+1$ varying between $1$ and 
$N$ and $\b=J-(n-1)D_{\rm x}$ varying from $1$ to $D_{\rm x}$.
We use these superindices to rewrite the terms in the square brackets 
as
\be
\sum_{m}v_{mqsj}G_{nm}\equiv \tilde{V}_{I,j}
\label{vd1}
\ee
for all $(qs)\in\callL$ and 
\be
\sum_{m}v_{mqjs}G_{nm}\equiv 
\tilde{V}^{\rm x}_{J,j}
\label{vx1}
\ee
for all $(qs)\in\callL_{\rm x}$.
In general, these quantities can be zero for several $j$'s since for a fixed 
pair $(qs)$ the values of $|v_{mqsj}|$ or $|v_{mqjs}|$ may be smaller 
than $\L$. To 
minimize the number of quantities to store we found convenient 
to define the sets
\bea
\callS(qs)&=&\{j:|v_{mqsj}|>\L\;{\rm for\;some}\;m\},
\\
\callS_{\rm x}(qs)&=&\{j:|v_{mqjs}|>\L\;{\rm for\;some}\;m\}.
\eea
Let $d(qs)$ and $d_{\rm x}(qs)$ be the dimension of $\callS(qs)$ 
and $\callS_{\rm x}(qs)$ respectively. For any given couple $(qs)$ we 
construct the map which associates to the integer $\s=1,\ldots,d(qs)$  the index 
$j(\s)\in \callS(qs)$ 
and to the integer  $\t=1,\ldots,d_{\rm x}(qs)$  the index 
$j(\t)\in \callS_{\rm x}(qs)$.  
Then, for every $j\in \callS(qs)$ we define 
\be
V_{I,\s}\equiv \tilde{V}_{I,j(\s)}
\ee
and for every $j\in \callS_{\rm x}(qs)$ we define 
\be
V^{\rm x}_{J,\t}\equiv \tilde{V}^{\rm x}_{J,j(\t)}.
\ee

We observe that for a fixed superindex $I=(\a,n)$ (or $J=(\b,n)$) the length of the 
array $V_{I,\s}$ (or $V^{\rm x}_{J,\t}$) depends on 
the pair $(q(\a),s(\a))\in\callL$ (or $(q(\b),s(\b))\in\callL_{\rm x}$). 
To calculate $V$ and $V^{\rm x}$ we define
two more sets
\bea
\callL(qs)&=&\{(mj):|v_{mqsj}|>\L\}
\\
\callL_{\rm x}(qs)&=&\{(mj):|v_{mqjs}|>\L\}
\eea
Let $D(qs)$ and $D_{\rm x}(qs)$ be the dimension of the sets 
$\callL(qs)$ and $\callL_{\rm x}(qs)$ respectively. Given $\m=1,\ldots D(qs)$ we 
establish a $(qs)$-dependent map to 
extract the couple $(mj)=(m(\m)j(\m))$. Similarly, given 
$\n=1,\ldots,D_{\rm x}(qs)$ we establish a $(qs)$-dependent map to 
extract the couple $(mj)=(m(\n)j(\n))$. We then construct two 
two-dimensional arrays 
\be
v_{\a\m}\equiv v_{mqsj},
\ee
with $(qs)=(q(\a)s(\a))\in\callL$ and 
$(mj)=(m(\m)j(\m))\in\callL(qs)$, and 
\be
v^{\rm x}_{\b\n}\equiv v_{mqjs},
\label{vxbetanu}
\ee
with $(qs)=(q(\b)s(\b))\in\callL_{\rm x}$ and 
$(mj)=(m(\n)j(\n))\in\callL_{\rm x}(qs)$. Notice that both $v$ and 
$v^{\rm x}$ have rows of different lenghts.
A code 
to fill up the array $V$ would have the following structure
\bea
{\tt DoLoop[I=1,D\times N]\{ }
\nn\\
{\tt \a(I),n(I) }
\nn\\
{\tt (qs)=(q(\a),s(\a))\in\callL }
\nn\\
{\tt DoLoop[\m=1,D(qs) ]\{ }
\nn\\
{\tt (mj)=(m(\m),j(\m))\in\callL(qs)}
\nn\\
{\tt V_{I,\s(j)}=V_{I,\s(j)}+v_{\a\m}G_{nm}}
\nn \\ \} \nn\\ \} \nn
\eea
Similarly for $V^{\rm x}$ we have
\bea
{\tt 
DoLoop[J=1,D_{x}\times N]\{ }
\nn\\
{\tt  \b(J),n(J)}
\nn\\
{\tt (qs)=(q(\b),s(\b))\in\callL_{x}}
\nn\\
{\tt DoLoop[\n=1,D_{x}(qs)]\{ }
\nn\\
{\tt (mj)= (m(\n),j(\n))\in\callL_{x}(qs)}
\nn\\
{\tt V^{x}_{J,\t(j)}=V^{x}_{J,\t(j)}+v_{\b\n}G_{nm}}
\nn \\ \} \nn\\ \} \nn
\eea

Once the arrays $V$ and $V^{\rm x}$ are filled the calculation of the 
self-energy is reduced to calculate
\be
\S_{ij}=2B_{ij}+X_{ij},
\label{2Bself2}
\ee
with the bubble term
\be
B_{ij(\s)}=\sum_{npr}\sum_{(qs)\in\callL}v_{npri}G_{pq}\bar{G}_{sr}V_{I,\s}
\label{bubble1}
\ee
(we recall that $I=[(qs)\in\callL,n$] and the 2-nd exchange term 
\be
X_{ij(\t)}=\sum_{npr}\sum_{(qs)\in\callL_{x}}v_{npri}G_{pq}\bar{G}_{sr}V^{\rm x}_{J,\t}
\label{2x1}
\ee
(we recall that $J=[(qs)\in\callL_{\rm x},n$].

Next we observe that for a fixed $i$ the sum over $n$ in $B$  and $X$ 
can be restricted to those $n$ for which the pair $(ni)\in\callL$ 
since $v_{npri}=v_{pnir}$. We then proceed as follows. 

- {\em Step 1)} For any fixed 
$(ni)\in\callL$ we construct the matrix
\be
H_{sp}^{(ni)}=\sum_{r}v_{npri}\bar{G}_{sr}
\ee
where the sum can be restricted to those $r$ for which the pair 
$(pr)\in\callL(in)$. In this way Eqs.~(\ref{bubble1},\ref{2x1}) 
becomes
\be
B_{ij(\s)}=\sum_{np}\sum_{(qs)\in\callL}G_{pq}H_{sp}^{(ni)}V_{I,\s}
\label{bubble2}
\ee
\be
X_{ij(\t)}=\sum_{np}\sum_{(qs)\in\callL_{\rm x}}G_{pq}H_{sp}^{(ni)}V^{\rm x}_{J,\t}
\label{2x2}
\ee

- {\em Step 2)} For any fixed $(qs)\in\callL$ we extract the superindex 
$I=[(qs),n]$ and construct the matrix
\be
Z_{i,I}=\sum_{p\in \callS(ni)}G_{pq}H_{sp}^{(ni)}.
\ee
Here the sum over $p$ is restricted to $\callS(ni)$ since 
$H_{sp}^{(ni)}$ vanishes for $p\notin \callS(ni)$. In terms of the 
$Z$ matrix the bubble term becomes
\be
B_{ij(\s)}=\sum_{I}Z_{i,I}V_{I,\s}
\label{bubble3}
\ee

- {\em Step 3)} For any fixed $(qs)\in\callL_{\rm x}$ we extract the superindex 
$J=[(qs),n]$ and construct the matrix
\be
Z^{\rm x}_{i,J}=\sum_{p\in \callS(ni)}G_{pq}H_{sp}^{(ni)}.
\ee
In terms of the 
$Z^{\rm x}$ matrix the 2-nd order exchange term becomes
\be
X_{ij(\t)}=\sum_{J}Z^{\rm x}_{i,J}V^{\rm x}_{J,\t}
\label{xtermfinal}
\ee

The structure of a code for the implementation of the above three 
steps would have the following structure
\bea
{\tt DoLoop[\a=1,D ]\{ }
\nn\\
{\tt (in)=(i(\a),n(\a))\in\callL}
\nn\\
{\tt DoLoop[\m=1,D(in)]}\{
\nn\\
{\tt (rp)= (r(\m),p(\m))\in\callL(in)}
\nn\\
{\tt DoLoop[s=1,N]}\{
\nn\\
{\tt H(s,p)=H(s,p)+v_{\a\m}*\bar{G}_{sr}}
\nn\\ \}  
\nn\\ \}
\nn\\
{\tt - Bubble -}
\nn\\
{\tt DoLoop[\a'=1,D]}\{
\nn\\
{\tt  (qs)=(q(\a'),s(\a'))\in\callL}
\nn\\
{\tt I=I[(qs),n]}
\nn\\
{\tt DoLoop[\g=1,d(ni)]}\{
\nn\\
{\tt  p=p(\g)\in\callS(in)}
\nn\\
{\tt Z=Z+G_{pq}*H(s,p)}
\nn\\ \}
\nn\\
{\tt DoLoop[\s=1,d(qs)]}\{
\nn\\
{\tt  j(\s)\in \callS(qs)}
\nn\\
{\tt B(i,j)=B(i,j)+Z*V_{I,\s}}
\nn\\ \}
\nn\\ \}
\nn\\
{\tt - Exchange -}
\nn\\
{\tt DoLoop[\b=1,D_{x}]}\{
\nn\\
{\tt (qs)=(q(\b),s(\b))\in\callL_{x}}
\nn\\
{\tt J=J[(qs),n]}
\nn\\
{\tt DoLoop[\g=1,d(ni)]}\{
\nn\\
{\tt  p=p(\g)\in\callS(ni)}
\nn\\
{\tt Z^{x}=Z^{x}+G_{pq}*H(s,p)}
\nn\\ \}
\nn\\
{\tt DoLoop[\t=1,d_{x}(qs)]}\{
\nn\\
{\tt  j(\t)\in\callS_{x}(qs)}
\nn\\
{\tt X(i,j)=X(i,j)+Z^{x}*V^{x}_{J,\t}}
\nn\\ \}
\nn\\ \}
\nn
\eea

We observe that the above implementation of the 2B self-energy 
requires at most two-dimensional arrays.  

\section{Discussion on scaling}
\label{scalingsec}

In this section we show that the scaling of the dissection 
algorithm with $N$ reduces with increasing number of vanishing Coulomb integrals. 
This means that it is possible to
take advantage of the sparse nature of the $v$ tensor 
without changing the implementation. Another advantage is that 
the convergence of a simulation
can easily be checked by reducing the 
cutoff, see also below. 

Let us first show that the algorithm scales like $N^{5}$ when all 
Coulomb integrals are larger than $\L$. In this case both sets $\callL$ 
and $\callL_{\rm x}$ have dimension $D=D_{\rm x}=N^{2}$ and, for any pair $(qs)$, 
the sets $\callL(qs)$ 
and $\callL_{\rm x}(qs)$ have dimension $D(qs)=D_{\rm 
x}(qs)=N^{2}$. Thus the calculation of $V$ and $V^{\rm x}$ involves an 
external loop of lenght $N^{3}$ and an internal loop of lenght 
$N^{2}$, resulting in a $N^{5}$ scaling. For the calculation of the 
bubble and the 2-nd order exchange terms we have an external loop of 
length $N^{2}$ followed by a cascade of two loops of lenghts $N^{2}$ 
and $N$ respectively, hence again a $N^{5}$ scaling, to build  
$H^{(ni)}_{sp}$. After closing 
these two loops, we have (for 
both $B$ and $X$)
the opening of a loop of lenght $N^{2}$  followed by a sequence of two loops of length $N$ 
since the dimension of the sets $\callS(qs)$ and $\callS_{\rm x}(qs)$ is 
$d(qs)=d_{\rm x}(qs)=N$ for all $(qs)$. We conclude that the overall 
scaling of the algorithm is $7N^{5}$.

The previous discussion helps in determining how the algorithm scales 
in the general case. Let
\be
M=\frac{1}{D}\sum_{(qs)\in \callL} D(qs),
\ee
\be
M_{\rm x}=\frac{1}{D_{\rm x}}\sum_{(qs)\in\callL_{\rm x}}D_{\rm x}(qs).
\ee
Then, the calculation of $V$ scales like $D\times N\times M$ whereas 
the calculation of $V^{\rm x}$ scales like $D\times N\times M_{\rm 
x}$, see the loop structure below Eq.~(\ref{vxbetanu}). 
From the loop structure below Eq.~(\ref{xtermfinal}) we infer that 
the calculation of $H^{(ni)}_{sp}$ scales like $V$. To determine the 
scaling of $B$ and $X$ we need to introduce two more integers
\be
m=\frac{1}{D}\sum_{(qs)\in \callL}d(qs),
\ee
\be
m_{\rm x}=\frac{1}{D_{\rm x}}\sum_{(qs)\in\callL_{\rm x}}d_{\rm x}(qs).
\ee
We then see that $B$ scales like $D^{2}\times 2m$ whereas $X$ scales 
like $D\times D_{\rm x}\times (m+m_{\rm x})$. 
We conclude that the dissection algorithm scales like 
\be
S_{\rm da}=D[N(2M+M_{\rm x})+2mD+(m+m_{\rm x})D_{\rm x}].
\label{scalingeq}
\ee
This number should be compared with the scaling $7N^{5}$ of an algorithm blind to the magnitude 
of the Coulomb integrals.

Let us consider a few examples. For an extended Hubbard interaction 
$v_{ijmn}\propto \d_{in}\d_{jm}$. In this case $D=N$, $D_{\rm x}=2N^{2}$, 
$M=N$, $M_{\rm x}=1$, $m=N$ and $m_{\rm x}=1$ and the overall 
calculations scales like $2N^{4}$. For a Hubbard like interaction 
$v_{ijmn}\propto \d_{in}\d_{jm}\d_{ij}$. In this case $D=N$, $D_{\rm 
x}=N$, $M=1$, $M_{\rm x}=1$, $m=1$ and $m_{\rm x}=1$ and the overall 
calculation scales like $7N^{2}$. More generally, the dissection 
algorithm minimize the scaling  by exploiting 
the sparsity of the four-index Coulomb tensor.
In Table \ref{scalingtab} we report the scaling for four different 
organic molecules (first column). The second column shows the number $N$  of 
bound Kohn-Sham states (either occupied or unoccupied) calculated 
using the Quantum Espresso package~\cite{QuantumEspresso} whereas  
the third column shows the number of electrons $N_{\rm 
el}$ per spin of the charge neutral molecule. We 
calculated the Coulomb integrals in the bound sector using the Yambo 
code~\cite{MARINI20091392} and performed convergence tests on the 
time-dependent charge density put in motion 
by a weak attosecond 
laser pulse using the CHEERS code~\cite{PS-cheers}. In all cases we 
found that the results converge by setting the cutoff $\L=0.01$ a.u. 
(the maximum value of the Coulomb integrals is about 0.5 a.u.). 
For all four molecules we calculated $D$, $D_{\rm x}$, $M$, $M_{\rm 
x}$, $m$  and $m_{\rm x}$ and report the scaling $S_{\rm da}$ of 
Eq.~(\ref{scalingeq}) in the fourth column. This should be compared 
with the scaling $7N^{5}$ in the fifth column. The gain factor 
$g\equiv 7N^{5}/S_{\rm da}$ is reported in the last column.

\begin{table}[b]
  \caption{Scaling of the dissection algorithm for four different 
  organic molecules. $N$ is the number of basis functions, $N_{\rm 
  el}$ is the number of electrons, $S_{\rm da}$ is the scaling of the 
  dissection algorithm, $7N^{5}$ is the scaling of an algorithm 
  blind to the magnitude of the Coulomb integrals and $g\equiv 7N^{5}/S_{\rm da}$ is the gain factor.}
  \begin{tabular}[htbp]{@{}cccccc@{}}
    \hline
    & $N$ & $N_{\rm el}$ & $S_{\rm da}$ & $7N^{5}$ & $g$\\
    \hline
    Glycine & 18 & 15  & $4.10\times 10^{6}$  & $1.32\times 10^{7}$ & 
    3.2 \\
    Phenylalanine & 36 & 32 & $7.56\times 10^{7}$  & $4.23\times 10^{8}$ & 5.6\\
    Tryptophan & 47 & 39  & $1.51\times 10^{8}$  & $1.61\times 10^{9}$& 
    10.6 \\
    Adenine & 57 & 25  & $1.19\times 10^{8}$  & $4.21\times 10^{9}$& 
    35.2 \\
    \hline
  \end{tabular}
  \label{scalingtab}
\end{table}

\section{Conclusions}
\label{conclusionsec}

We have described an algorithm to calculate the 2B self-energy 
appearing in the collision integral of the KBE and the GKBA equation. The basic idea 
consists in dissecting the set of Coulomb integrals in properly 
chosen (overlapping) subsets to exploit the  sparsity of 
the Coulomb tensor. We have shown that the scaling of the 
computational cost reduces to $N^{4}$ for density-density type 
interactions and to $N^{2}$ for Hubbard-like interactions. 
For systems like atoms and molecules the net gain with respect to a 
$N^{5}$ scaling depends on the nature of the atoms. Our empirical 
evidence  is that the gain factor increases with increasing 
the ratio $N/N_{\rm el}$, see Table~\ref{scalingtab}.

The dissection algorithm as currently implemented in CHEERS~\cite{PS-cheers} 
is further optimized by exploiting the symmetries in 
Eq.~(\ref{symprop}). This leads to a reduction by a factor of 2 of 
the dimension of the sets  $\callL$ and $\callL(qs)$. The entire 
procedure can easily be generalized to complex (already implemented 
in CHEERS) and spin-dependent basis functions.

\begin{acknowledgement}
 G.S. and E.P.  acknowledge EC funding through the RISE Co-ExAN (Grant No. GA644076).
E.P. also acknowledges funding from the European Union project
MaX Materials design at the eXascale H2020-EINFRA-2015-1, Grant Agreement No.
676598 and Nanoscience Foundries and
Fine Analysis-Europe H2020-INFRAIA-2014-2015, Grant Agreement No. 654360.
G.S. also 
acknowledge Tor Vergata University for financial support through the Mission Sustainability Project
2DUTOPI.
\end{acknowledgement}

%
\bibliographystyle{pss}
%


\providecommand{\WileyBibTextsc}{}
\let\textsc\WileyBibTextsc
\providecommand{\othercit}{}
\providecommand{\jr}[1]{#1}
\providecommand{\etal}{~et~al.}

%

\end{document}